\documentclass[twocolumn,showpacs,showkeys,amsmath,amssymb,floatfix]{revtex4}

\usepackage{graphicx}
\usepackage{subfigure}
\usepackage{color}

\begin{document}

\title{Influence of the vacuum interface on the charge distribution 
       in V$_2$O$_3$ thin films}

\author{U.~Schwingenschl\"ogl$^1$, R.~Fr\'esard$^2$, and V.~Eyert$^3$}

\affiliation{$^1$KAUST, PCSE Division, P.O. Box 55455, Jeddah 21534, Saudi Arabia\\
$^2$Laboratoire CRISMAT, UMR CNRS-ENSICAEN(ISMRA) 6508, and IRMA, FR3095, Caen, France\\
$^3$Center for Electronic Correlations and Magnetism, Institute for Physics, University
of Augsburg, 86135 Augsburg, Germany}

\date{\today}

\begin{abstract}
The electronic structure of V$_2$O$_3$ thin films is studied by means of the
augmented spherical wave method as based on density functional theory and 
the local density approximation. We establish that effects of charge 
redistribution, induced by the vacuum interface, in such films are 
restricted to a very narrow surface layer of $\approx$ 15\,\AA\ thickness. 
As a consequence, charge redistribution can be ruled out as a source of 
the extraordinary thickness-dependence of the metal-insulator transition 
observed in V$_2$O$_3$ thin films of $\sim$ 100-1000\,\AA\ thickness.
\end{abstract}


\maketitle

\section{Introduction}

Over the last decades, V$_2$O$_3$ has been investigated extensively by 
both experiment and theory -- and still is a topic of great interest 
(see, e.g., Refs.\ \cite{rodolakis09,borghi09} and the literature given 
therein). Today, V$_2$O$_3$ thin films are receiving growing attention,
particularly because the thin film geometry appears to strongly modify 
the electronic properties of the material as is most clearly demonstrated 
by the famous metal-insulator transition observed in bulk V$_2$O$_3$ as 
a function of temperature, pressure or doping \cite{imada98}. It is 
known that even a small alteration of the 
crystal structure, by Cr substitution \cite{kuwamoto80}
or application of external pressure \cite{limelette03}, can have a serious
effect on the metal-insulator transition in the bulk material. Therefore, 
it would not at all be surprising if structural modifications induced 
by the thin film geometry had similar severe implications.

Recently, the observation of a thickness-dependent metal-insulator 
transition in ultrathin V$_2$O$_3$ films was attributed to the 
increasing $c$/$a$ lattice parameter ratio (decreasing $a$, 
increasing $c$) because of interaction with the substrate \cite{luo04}. 
However, this conjecture has been refuted by both x-ray diffraction 
studies for high quality films \cite{grygiel07,allimi08} and density 
functional calculations \cite{sfe}. In fact, the experiments point 
to a very complex thickness-dependence of the lattice parameters 
with unsystematic changes of the $c$ lattice constant for film 
thicknesses of $\sim$ 100--1000\,\AA, thereby calling for a more 
in-depth study of the electronic properties of these films. Indeed, 
in other systems, like the LaAlO$_3$/SrTiO$_3$ heterostructure 
\cite{CPL467}, it has been shown that interfaces can seriously 
modify the electron density profile. A strong dependence of the 
electronic properties on the layer thickness has been also reported 
for the hexagonal layer compounds such as $ {\rm WS_2} $ \cite{klein2001}. 
In the present case a systematic investigation is still missing.

In this paper, we aim at clarifying the influence of the vacuum interface
on the electronic states of V$_2$O$_3$ thin films. At this interface, 
two possible scenarii may cause a significant reorganization of the 
electronic states:  (1) a local modification of the crystal structure
and  (2) a charge redistribution due to the broken symmetry at the 
vacuum interface. 
In general, charge redistribution due to a surface, similar to 
heterointerfaces, may induce a considerable deviation from the bulk 
properties, see, e.g., Refs.\ \cite{surf2,cpl449,gemming06,epl08}.
Even though possibility (2) therefore seems to be promising, it is 
the aim of this paper to point at the limitations of such a mechanism 
for film thicknesses exceeding some 15\,\AA.

\section{Computational Method and Technical details}

The calculations are based on density-functional theory and the local 
density approximation (LDA). They were performed using the 
scalar-relativistic implementation of the standard augmented spherical 
wave (ASW) method (see Ref.\ \onlinecite{aswrev}, Chap.\ 2 of Ref.\ 
\onlinecite{aswbook} and references therein).
In the ASW method, the wave function is expanded in atom-centered
augmented spherical waves, which are Hankel functions and numerical
solutions of Schr\"odinger's equation, respectively, outside and inside
the so-called augmentation spheres. In order to optimize the basis set,
additional augmented spherical waves were placed at carefully selected
interstitial sites. The choice of these sites as well as the augmentation
radii were automatically determined using the sphere-geometry optimization
algorithm \cite{sgo}. Self-consistency was achieved by a highly efficient
algorithm for convergence acceleration \cite{mixpap}. The Brillouin zone
integrations were performed using the linear tetrahedron method with an 
increasing number of {\bf k}-points  in order to check the convergence 
with respect to the granularity of the {\bf k}-point grid
\cite{bloechl94,aswbook}.

\begin{figure}[t]
\centering 
\includegraphics[width=0.35\textwidth,clip]{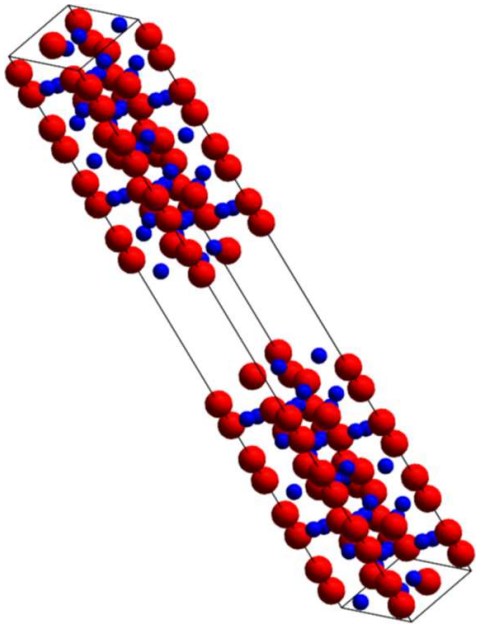}
\caption{(Color online) Slab used for modeling the V$_2$O$_3$ surface.
Large and small spheres represent the V and O atoms, respectively.} 
\label{fig0}
\end{figure}

The V$_2$O$_3$ surface was simulated by means of a fivefold superstructure
of the canonical hexagonal unit cell. The latter comprises six formula units,
i.e.\ six V layers along the (hexagonal) $c$ axis. These layers are separated
by O layers, giving rise to VO$_6$ octahedra. For further details see, 
e.g., Refs.\ \cite{us0304,eyert05b} and the references therein. The lattice
constants and fractional coordinates of the atoms were taken from Ref.\ 
\cite{dernier70}. The surface was then generated by removing all atoms 
from one of the five hexagonal unit cells, see Fig.\ \ref{fig0}, such that
one of the resulting surfaces terminates with an O layer.
This vacuum interface is the subject of the subsequent discussion.

In order to separate the charge redistribution due to the broken
crystal symmetry and the changes in chemical bonding at the vacuum 
interface from the structural relaxation effects at the V$_2$O$_3$ 
surface we will not include the latter in our considerations. In 
fact, structural relaxation is restricted to a very narrow range of  
about five atomic layers as has been pointed out in a detailed study  
by Kresse {\it et al.}\ \cite{kresse04}, which also provides an 
elaborate overview over the relevant literature. In contrast, we 
shall focus on the question whether the charge redistribution coming 
with the removal of bonding partners could affect the behavior of a 
much thicker slab of several 100--1000\,\AA. Consequently, accounting 
for the near-surface structural modification is, on the one side, 
dispensable and, on the other side, disadvantageous for a systematic 
analysis of the decay of charge redistribution effects.

Of course, in view of the strong electronic correlations present in 
$ {\rm V_2O_3} $, calculations beyond the standard LDA seem necessary. 
Indeed, while LDA fails to capture the metal-insulator transition 
of the bulk material, the latter could be successfully described using 
LDA+DMFT \cite{held01,keller04}. However, for the large supercell used 
in the present simulation of the surface the rather complex LDA+DMFT 
calculations would be beyond present day computational facilities. 
Even more important, the previous LDA+DMFT studies revealed that the 
strong electronic correlations come into play via the narrowing of the 
bands as calculated within the LDA, which occurs on going from the 
crystal structure of the metal to that of the insulator. Hence, there 
is a clear signature already in the LDA bands indicating the 
metal-insulator transition. As a consequence, although LDA has 
turned out to be insufficient in describing the metal-insulator 
transition itself, it clearly signals its occurrence. For that reason, 
it is well justified to omit a full LDA+DMFT treatment as long as the 
LDA results are carefully interpreted.

\section{Results}

\begin{figure}[b]
\centering 
\includegraphics[width=0.45\textwidth,clip]{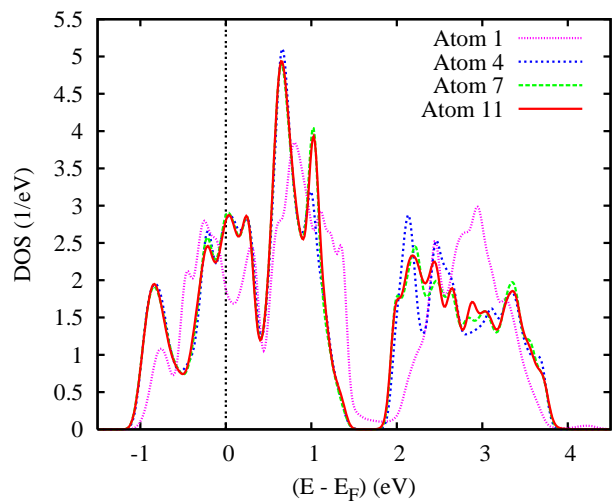}
\caption{(Color online) Site-projected partial V $3d$ DOS for a series 
          of V sites in a V$_2$O$_3$ thin film. Numbering of the V atoms 
          corresponds to their distance from the surface.} 
\label{fig1}
\end{figure}

\begin{figure*}[t]
\centering
\subfigure[]{\includegraphics[width=0.45\textwidth,clip]{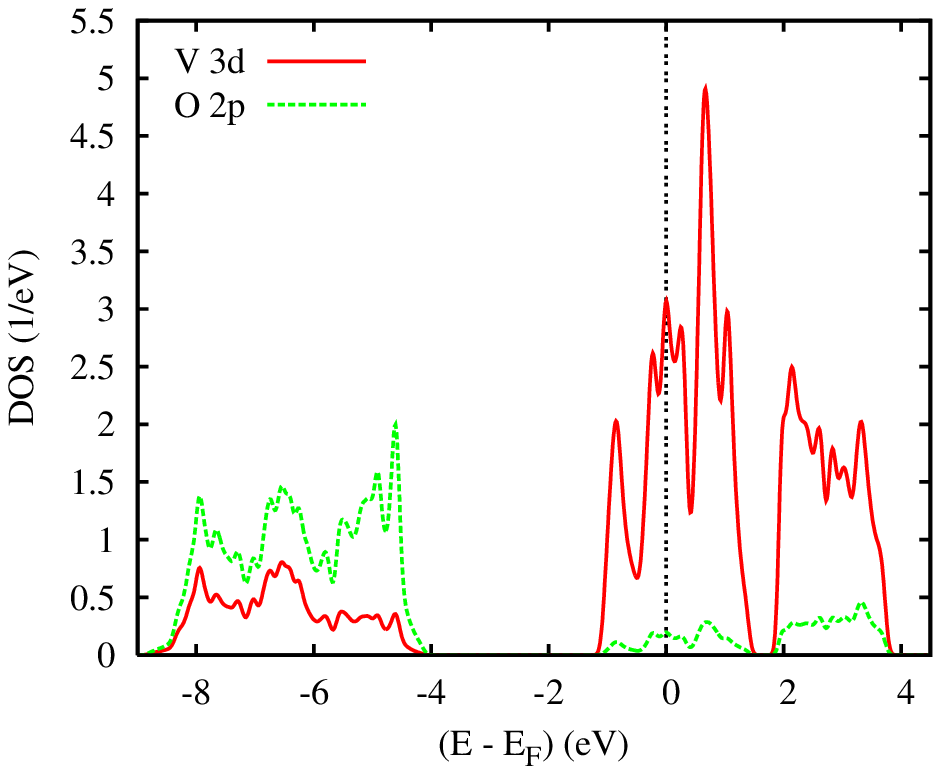}}
\hspace{0.05\textwidth}
\subfigure[]{\includegraphics[width=0.45\textwidth,clip]{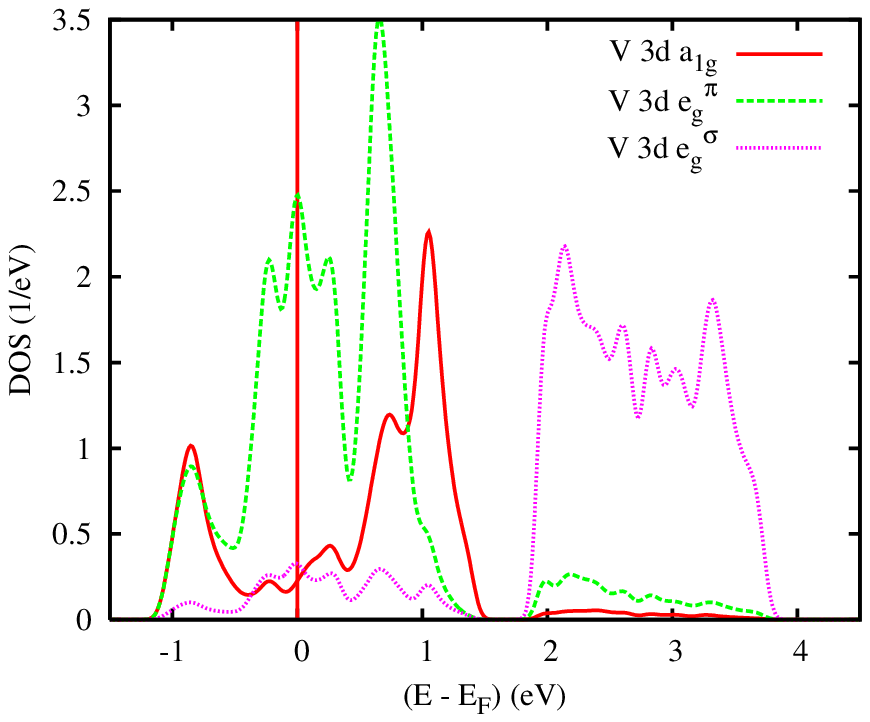}}
\\[0.1cm]
\subfigure[]{\includegraphics[width=0.45\textwidth,clip]{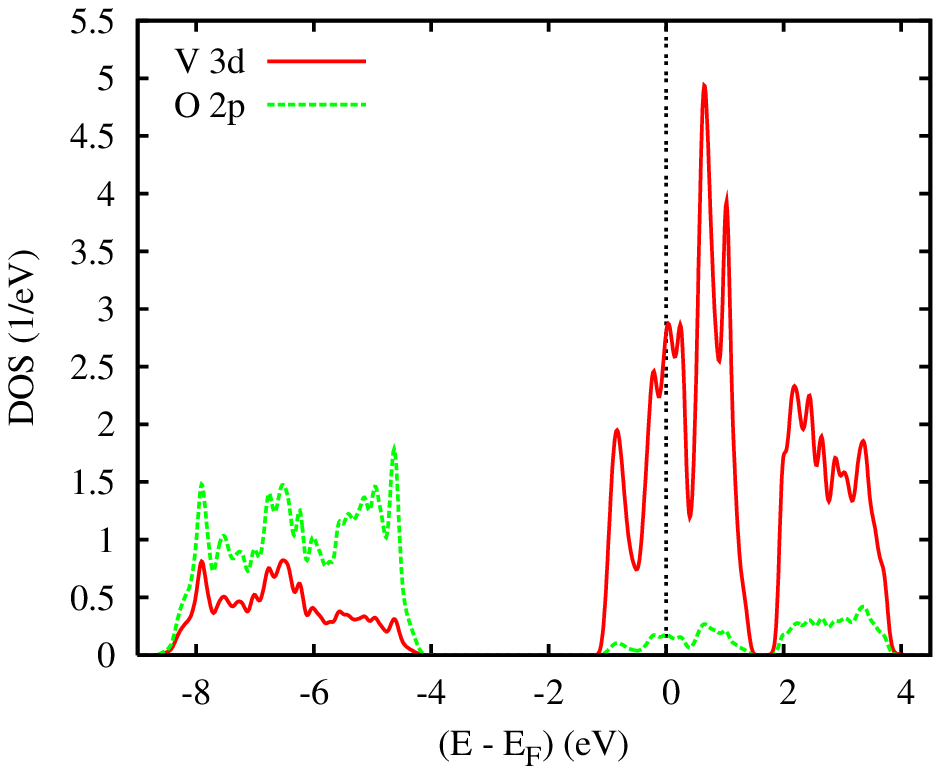}}
\hspace{0.05\textwidth}
\subfigure[]{\includegraphics[width=0.45\textwidth,clip]{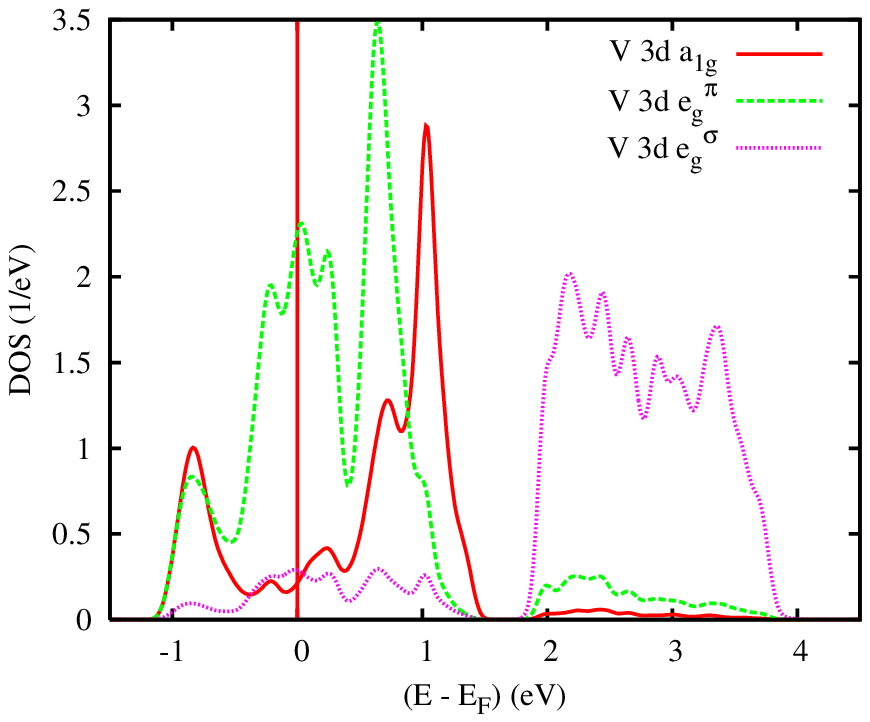}}
\\[0.1cm]
\subfigure[]{\includegraphics[width=0.45\textwidth,clip]{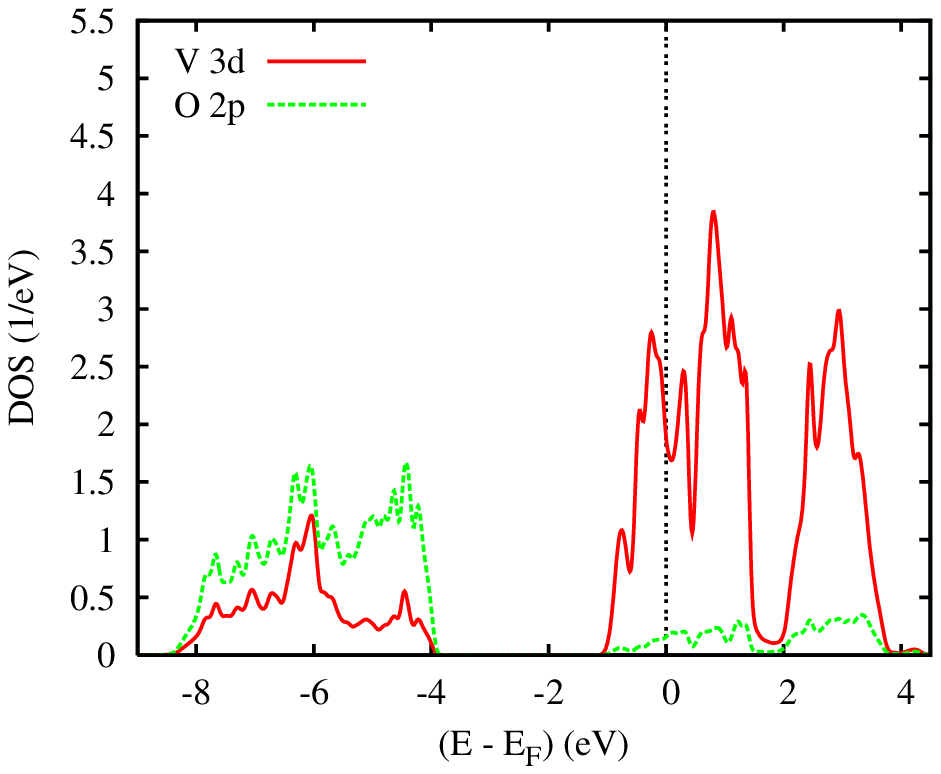}}
\hspace{0.05\textwidth}
\subfigure[]{\includegraphics[width=0.45\textwidth,clip]{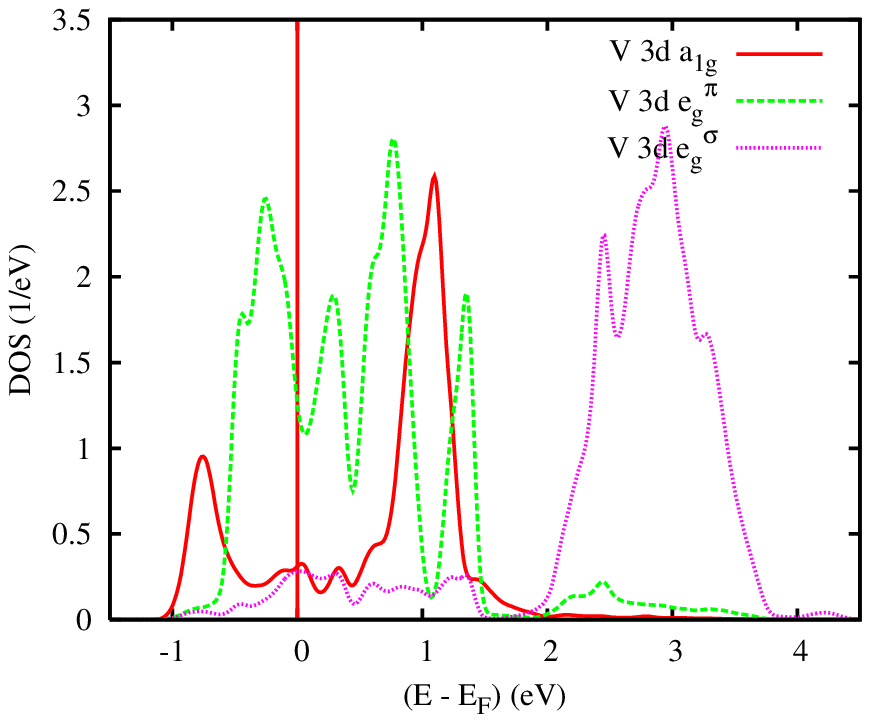}}
\caption{(Color online) Partial V $3d$ and O $2p$ DOS (left) as well as 
         V $3d$ symmetry components (right) for a VO$_6$-octahedron in 
         bulk V$_2$O$_3$ (top row), separated by $\approx 28$\,\AA\ from 
         the surface of a V$_2$O$_3$ thin film (center row), and at the 
         surface of a V$_2$O$_3$ thin film (bottom row). All DOS curves 
         are per formula unit.}
\label{fig2}
\end{figure*}

\begin{figure*}[t]
\centering
\hspace*{-0.07\textwidth}
\subfigure[]{\includegraphics[width=0.45\textwidth,clip]{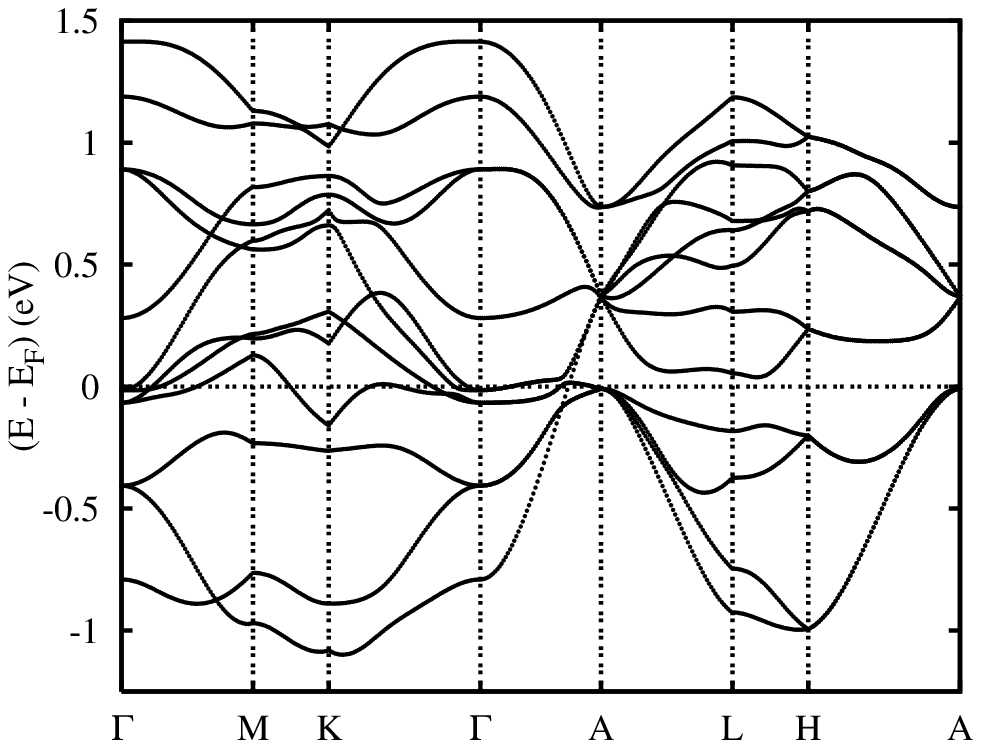}}
\hspace*{0.18\textwidth}
\subfigure[]{\includegraphics[width=0.25\textwidth,clip]{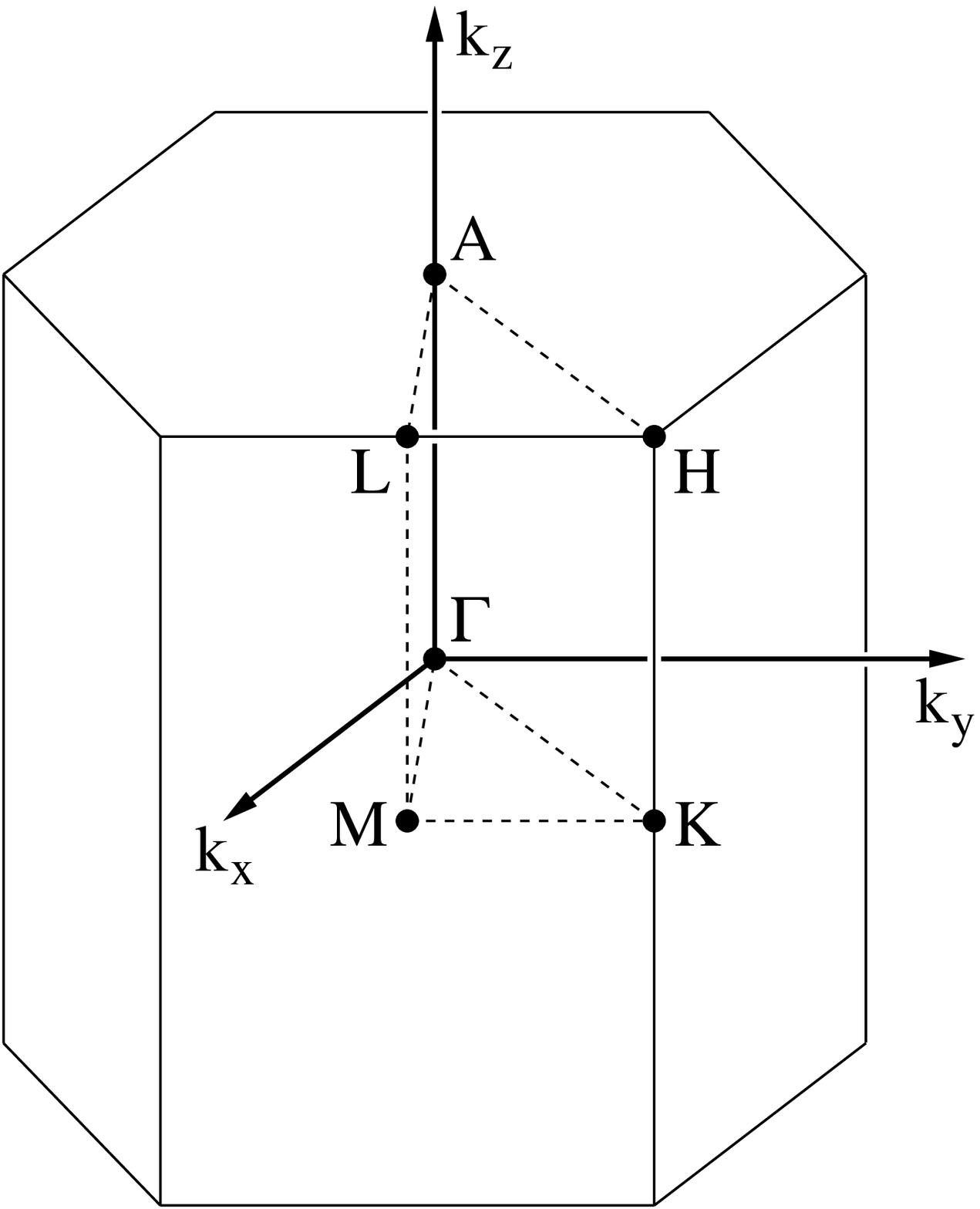}}
\\[0.2cm]
\subfigure[]{\includegraphics[width=0.45\textwidth,clip]{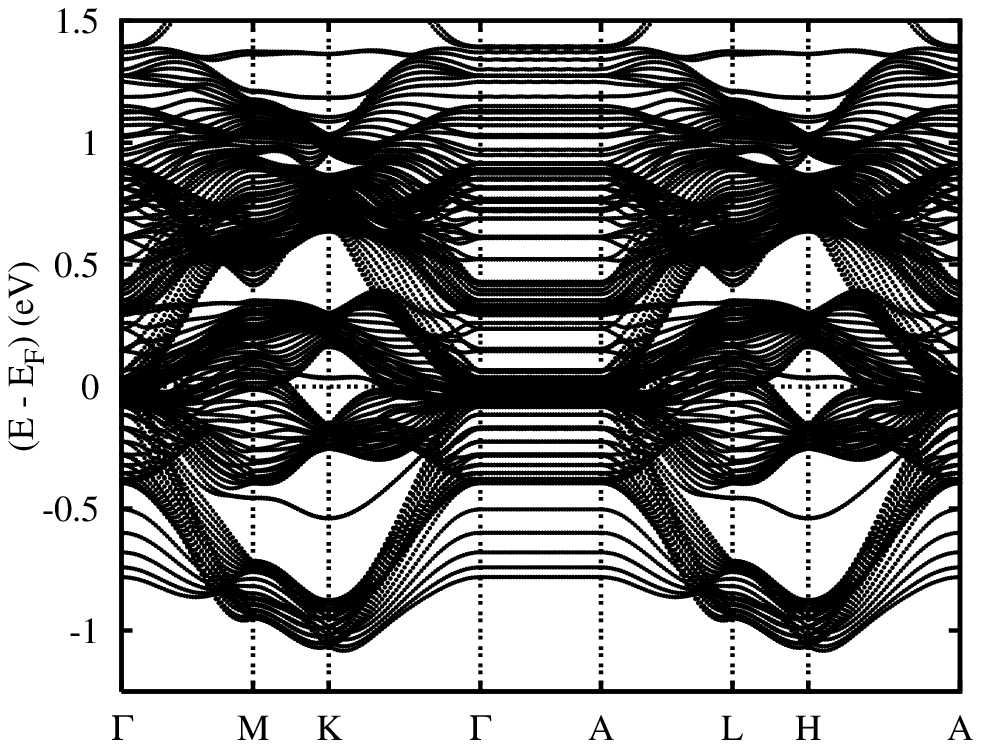}}
\hspace{0.05\textwidth}
\subfigure[]{\includegraphics[width=0.45\textwidth,clip]{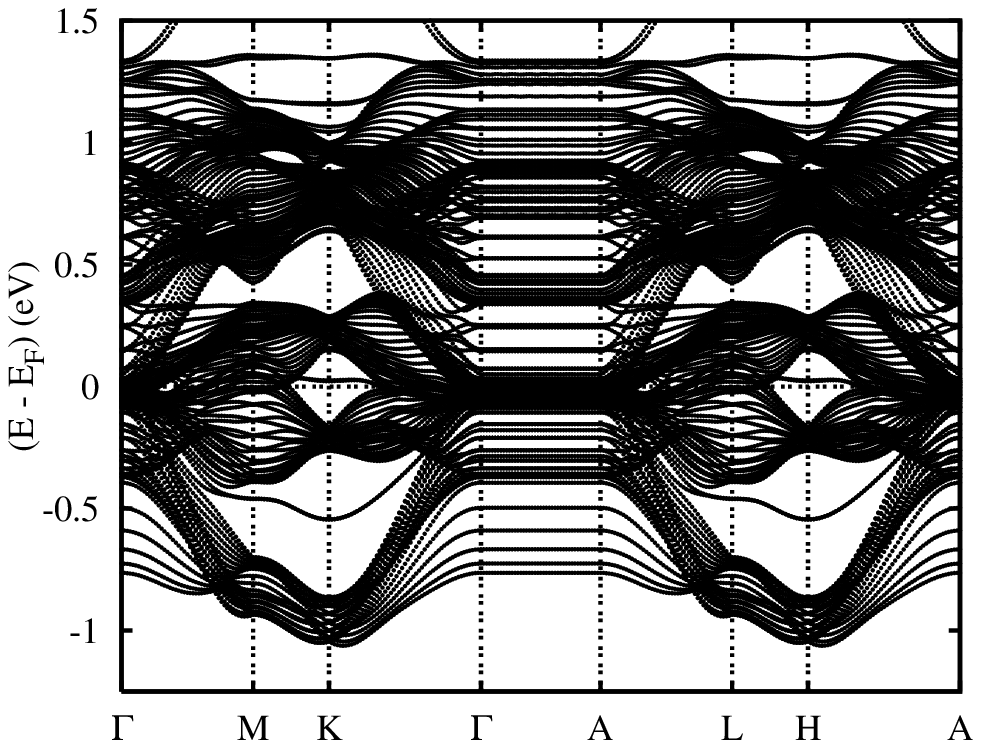}}
\caption{Electronic band structure of bulk V$_2$O$_3$ (top left),
         for high-symmetry lines in the first Brillouin zone of 
         the hexagonal lattice (top right), compared to a V$_2$O$_3$ 
         thin film with bulk lattice constants (bottom left) and 
         with an 1\% elongated $c$ lattice constant (bottom right).}
\label{fig3}
\end{figure*}

Due to the octahedral O-coordination of the V atoms in V$_2$O$_3$, the 
V $3d$ electronic bands (located around the Fermi level) are subject to 
a splitting in two subgroups. While the partially occupied t$_{2g}$ 
manifold is found in the energy range from $-1.2$\,eV to 1.4\,eV, 
the unoccupied $e_g^\sigma$ states extend from $1.7$\,eV to 3.7\,eV. 
These two groups are clearly observed in Fig.\ \ref{fig1}, 
which depicts the site-projected V $3d$ density of states (DOS) as 
obtained for various atoms in the thin film. Atoms are numbered 
according to their distance from the surface with atom 1 being 
located directly at the surface. We find that the shapes of the curves in 
Fig.\ \ref{fig1} are closely related to each other. Only for atom 
1, which is subject to the strongest interface effects, deviations
are visible since half of the bonding partners are missing and the
V--O therefore is reduced. In fact, the DOS curves of atoms 7 and 11,
which are separated from the surface
by about 18\,\AA\ and 28\,\AA, respectively, virtually resemble the
bulk DOS shape. For this reason, the decay of the interface-induced
relaxation of the electronic states is found to be very fast.

In order to establish further insight into the above-mentioned 
decay, we display in Fig.\ \ref{fig2} 
V $3d$ and O $2p$ partial DOSs in a wider energy range in the left 
column as well as orbital resolved V $3d$ partial DOSs in the right 
column. The nearly perfect splitting of the t$_{2g}$ manifold into 
$a_{1g}$ and $e_g^\pi$ sub-groups traces back to the trigonal
symmetry of the corundum lattice \cite{held01}. The partial DOS 
of V$_2$O$_3$ bulk (top row of Fig.\ \ref{fig2}) is almost identical 
to that found for a VO$_6$ octahedron in the thin film some 
28\,\AA\ beneath the surface (center row of Fig.\ \ref{fig2}). This 
holds for both the O $2p$ and the V $3d$ states. In contrast, the 
surface VO$_6$ octahedron clearly shows distinct electronic 
relaxation effects of the local electronic states (bottom row of 
Fig.\ \ref{fig2}), which can be attributed to the modified chemical 
$ d $-$ p $ bonding at the surface. In particular, for those V and
O atoms, which miss a bonding partner due to the surface, we 
observe a reduction of the $3d$ and $2p$ occupation, respectively, 
of approximately 0.08 electrons. We point out again that in obtaining 
these results we did not take into account the V$_2$O$_3$ surface 
relaxation in order to distinguish between the electronic changes 
and the short-range structural distortions. Our results entail that 
the effect of the vacuum interface on the local electronic states,
as reflected by the site-projected DOS, vanishes at about 15\,\AA\ 
below the surface. Charge redistribution induced by the surface is 
therefore restricted to the same narrow surface layer, in which the 
structural relaxation is observed \cite{kresse04}.

In a next step we turn to the discussion of $ {\bf k} $-resolved 
quantities and display the electronic bands in Fig.\ \ref{fig3}. 
The top row of Fig.\ \ref{fig3} depicts the band structure of bulk 
V$_2$O$_3$ as well as the non-primitive hexagonal Brillouin zone of 
the corundum lattice of V$_2$O$_3$, which here and in the following 
is used for representing findings of band structure calculations. 
The bulk results serve as a reference for comparison with the
surface results given in the bottom left panel of Fig.\ \ref{fig3}. 

Of course, the surface band structure comprises a lot more bands due 
to the simulation of the surface by the fivefold supercell of the 
surface slab as mentioned above. In this supercell, the increased 
number of different bands reflects the larger number of inequivalent 
atomic sites. V (and O) atoms with different distances to the vacuum 
are no longer equivalent from the crystallographic point of view. 
Furthermore, the dispersion along the high symmetry line $\Gamma$--A 
is strongly suppressed due to the fact that this line corresponds to 
the direction perpendicular to the surface. Note that the construction 
of the surface slab in a fivefold supercell of the hexagonal V$_2$O$_3$ 
cell implies that the length of the line $\Gamma$--A is reduced to a 
fifth and that the electronic bands are folded back. In order to enable
a straightforward comparison with the bulk data, the line $ \Gamma$-A 
is thus artificially stretched to the original length in the slab band 
structure. Having this fact in mind, we eventually find that the 
surface slab reproduces the characteristics of the bulk V$_2$O$_3$ 
bands.  In particular, there is no sign of an additional state, not
present in the bulk, which could be responsible for the anomalous 
behavior of thin films.

It has been found for V$_2$O$_3$ thin films, that the $a$ and $c$ 
constants of the corundum lattice deviate slightly, i.e.\ less than 
1\%, from their bulk values \cite{grygiel07,allimi08}. In addition, 
we have shown that the influence of the $a$ parameter change on the 
electronic structure is negligible \cite{sfe}. Here, we investigate 
whether modifications of the $c$ parameter may play a more important 
role by performing a surface slab calculation with the $c$ constant 
increased by 1\% as compared to the bulk. The results are 
depicted in the bottom right panel of Fig.\ \ref{fig3}, which uses 
the same representation as before. Comparison with the bottom left 
panel of Fig.\ \ref{fig3} shows that an elongation of the $c$ axis 
does not lead to a qualitative change of the electronic states. The 
two band structures are, in fact, almost identical. Hence, we obtain 
that the experimentally observed alterations of the  V$_2$O$_3$ lattice 
parameters in a thin film influence the electronic states only marginally, 
as long as they are not accompanied by a relaxation of the atomic 
positions. Indeed, a substantial lattice relaxation has been advocated 
to interpret transport measurements \cite{grygiel08b}.

\begin{figure}[htb]
\centering        
\includegraphics[width=0.45\textwidth,clip]{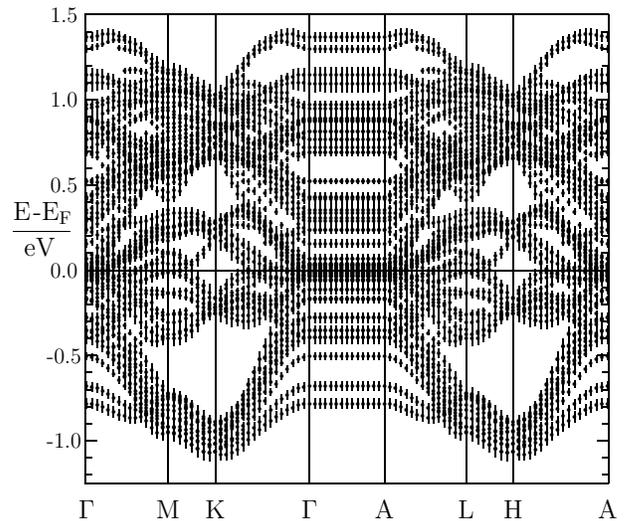}
\caption{Weighted electronic bands for a prototypical V atom in a 
         V$_2$O$_3$ thin film, located $\approx 28$\,\AA\ beneath 
         the surface. The length of the bars shown for each state 
         is proportional to the contributions of this atom. States
         with a contribution below a certain threshold are omitted 
         for clarity.}
\label{fig4}
\end{figure}

To ensure that the thin film electronic structure fully resembles the 
bulk a final step is necessary. We still have to exclude differences 
in the contribution of the atomic orbitals to the different electronic 
states, since they would point at a modified chemical bonding. Orbitally 
weighted band structures are useful tools to address this question.
Fig.\ \ref{fig4}
gives an example of a prototypical V atom, some 28\,\AA\ below the 
surface. Very similar results are obtained for the neighboring V 
sites. In Fig.\ \ref{fig4}, the bars added to each state have a length 
proportional to the contribution of the $3d$ $t_{2g}$ orbitals of 
the atom under consideration. States with minor $3d$ $t_{2g}$ 
admixture are not depicted for clarity. For this reason, some 
of the bands shown in the bottom left panel of Fig.\ \ref{fig3} 
are missing in Fig.\ \ref{fig4}. We observe that the selected 
V site contributes to the $3d$ $t_{2g}$ bands in the entire 
energy range from $-1.2$\,eV to 1.4\,eV and thus perfectly 
reflects the behavior of a bulk atom, which is true for any
atom located at least $\sim$ 15\,\AA\ below the surface. Charge 
redistribution as induced by the vacuum interface therefore has to be 
excluded as a possible source for the novel electronic properties 
of V$_2$O$_3$ thin films. We finally mention that in the case of
a heterosystems, like LaAlO$_3$/SrTiO$_3$ \cite{pentcheva08,lao},
different charge redistributions have been obtained in LDA and LDA+U
calculations. However, for the present system we do not find such a dependence.

In conclusion, our findings entail that neither the observed changes 
of the $a$ and $c$ lattice parameters in V$_2$O$_3$ thin films nor 
the response of the electronic system to the presence of the vacuum 
interface can explain significant deviations from the electronic 
properties of bulk V$_2$O$_3$. Only next to the surface, i.e.\ in a 
narrow layer of at most $15$\,\AA\ thickness, a charge redistribution 
is observed. The same behavior has been found in calculations
for other transition metal surfaces \cite{fu84,lee86}.

Below the $15$\,\AA\ surface layer the electronic states are hardly 
modified by the vacuum interface. LDA+DMFT calculations have shown 
that the narrowing of the V $3d$ $t_{2g}$ group of bands seen in the 
plain LDA DOS (which traces back to structural alterations 
\cite{eyert05a}) is the essential criterion for obtaining a 
metal-insulator transition in bulk V$_2$O$_3$ \cite{held01,keller04}. 
Because such a narrowing is not observed in our results, they 
contradict a transition for V$_2$O$_3$ thin films of 100-1000\,\AA\ 
thickness, as long as a restriction of the structure relaxation to 
the surface layer is assumed. Reversely, only further changes of the 
atomic structure, probably in the entire thin film, can explain the 
experimental situation. A detailed study of the crystal structure 
over the entire thin film is thus required to settle this issue.

\section*{Acknowledgment}

The work was supported by the Deutsche Forschungsgemeinschaft  through 
SFB 484.

\end{document}